# PROVIDE A MODEL FOR HANDOVER TECHNOLOGY IN WIRELESS NETWORKS

Abbas Asosheh[1], Nafiseh Karimi[2] and Hourieh Khodkari[3]

[1]Faculty of Technical Engineering, Tarbiat Modares University, Tehran, Iran
asosheh@modares.ac.ir
[2] Faculty of Technical Engineering, Tarbiat Modares University, Tehran, Iran
n.karimi@modares.ac.ir
[3] Faculty of Technical Engineering, Tarbiat Modares University, Tehran, Iran
khodkari@gmail.com

*ABSTRACT*

*Fast Handovers for the MIPv6 (FMIPv6) has been proposed to reduce the Handover latency, in the IETF. It could not find the acceptable reduction, so led to more efforts to improve it and however the creation of multiple Handover methods in the literature.*

*A stable connection is very important in mobile services so the mobility of device would not cause any interruption in network services and thus mobility management plays a very important role. Mobile IPv6 has become a general solution for supporting mobility between different networks on the internet which a flawless connection needs to be managed properly.*

*In order to select the appropriate method¤ in this paper, all the proposed methods have been classified according to the identified performance metrics.* Call blocking probability, *Handover blocking probability, Probability of an unnecessary handover, Duration of interruption* and delay, as the most important Handover algorithm *performance metrics are* introduced.

*The AHP method will be deployed to weight the metrics in a sample topology according to the selected sound application. Then the TOPSIS method will be employed to find the appropriate Handover algorithm.*

## 1    Introduction

IPv6 is a next generation network protocol, which was standardized to take the place of current protocols. This protocol will become the infrastructure of the next generation internet and in comparison with IPv4, it has improved dramatically in these areas: security, dynamism, convergence, scalability and was standardized in 1990s by IETF.[1] Integrated management in next generation network provides management functions for NGN resources and maintains connections between management plans themselves and other NGN renounces or services.[1] MIPv6 is seen as the de facto standard for mobility management in next  generation networks (NGN) with IPv6 nodes.[2]

A management framework is needed in order to improve the costumer service satisfaction and simultaneously decrease the operator expenses using new technology, business models and new functional methods. One of the available services included in next generation networks is the possibility of communication between different devices and connections among fixed networks and mobile ones or wired and wireless networks. Such service requires a secure and reliable environment and to gain more efficient results it must be used with a proper management framework.[1]

The handover process happens when the MN(Mobile Node) moves from one access medium to another, and it should accomplish three operations: movement detection, new CoA(Care-of Address) configuration, and BU(Binding Update).[3] To make a MN stay connected to the Internet regardless of its location, mobile IPv6 is proposed as the next generation wireless Internet protocol. This is achieved primarily through using CoA to indicate the location of the MN. Although the Mobile IPv6 protocol has many promising characteristics and presents an elegant mechanism to support mobility, it has an inherent drawback. That is, during a handover process, there is a short period that the mobile node is unable to send or receive packets because of link switching delay and IP protocol operations.[4] This handover delay is intolerable for most applications. Proposed methods, mostly with study on most effective parameters in improving the QoS(Quality of Service) , including improve delay ,jitter and packet lost parameters are trying to improve the performance of Handover. But regardless of categories, in different conditions, the proposed methods will not enough performance, and a pretreatment is necessary to distribute the criteria in various classes having the same characteristics e.g. delay and jitter.[5]

A stable connection is very important in a mobile network so the mobility of device would not cause any interruption in network services. It shows the importance of the mobility management role. To determine the parameters that affect the performance of handover, classification of existing methods is required. It is also necessary to determine handling handover procedures. After identifying the parameters that can affect the efficiency of handover, choosing the appropriate algorithm can be done by using Multi-Criteria Decision Making Methods.

When looking on a handover from an architectural point of view there are two different types, vertical and horizontal. The horizontal handover is a handover between base stations belonging to the same type of network technology while the vertical handover is made between base stations attached to different network technologies.[6] MIH framework is a standard being developed by IEEE802.21 which proposes to enable handover between heterogeneous networks.[7]

From the perspective of geographical, mobility management solutions are divided in to two categories: macro-mobility and micro-mobility solutions. The mobility between two network domains known as macro-mobility and between the subnets in a domain known as micro-mobility. Several micro protocols have been proposed, which include HAWAII (Handover-Aware Wireless Access Internet Infrastructure)[8], CIP(Cellular IP)[9], HMIP (Hierarchical MIP)[10], IDMP (Intra-Domain Mobility Management).[11]

Due to the time of connection to new access point and its better management, three types of Handover are defined. In the hard handover scheme the MN changes its point of attachment with a short interruption of service. The old link is released and a new one created at the new BSs. The time the system needs to set up the path is referred to as the network response time. If the old radio link is broken up before the network completes the setup, the connection is dropped even if there are channels available in the cell.[12] Therefore this method is called brake before make.[13]

The seamless handover is based on the concept of changing between cells using the old and the new connection simultaneously with only one of them being active. Data is broadcast via both links. The old link stays active as long as the new path is activated. In comparison to the hard handover the seamless approach is more reliable since the old link is release after a new one has been established. However the utilization of two links during the handover phase degrades the number of available channels, which has a negative impact on the number of users that can be carried.[12]

The soft handover allows a transient phase during which multiple links can be used for communication simultaneously with all of them being active - which has the advantage that if one link fails the MN can communicate using the remaining links -. Soft handover can be used to extend the time that is available to make a handover decision without any loss of QoS. This allows reduction of the service interruption to a minimum when changing between cells. However in addition to limiting the efficient use of the frequency spectrum, this results in high data overhead since packets are transmitted on all links.[12]

When looking on a handover from layer point of view there are different types, The sub network layer, network layer, transport layer, session layer and application layer, that the SCTP(Stream Control Transmission Protocol), SLM(Session Layer Mobility Management) and SIP(Session Initiation Protocol) Handover procedures are examples of transport, sessions and application layer, respectively.[14],[15]

In the literature, handover performance metrics in order to select handover algorithm is as follows**:** Call blocking probability, Handover blocking probability, Handover probability, Call dropping probability, Probability of an unnecessary handover, Rate of handover, Duration of and Delay**.**[16],[17]

A number of procedures for handling handoffs have been proposed in the literature**.** A common handoff priority scheme is one in which a specified number of channels is set aside for the exclusive use of handoffs. The number to be set aside can be made adaptable with traffic intensity to satisfy a given handoff dropping/blocking probability combination. This priority strategy is often termed a guard-channel approach. Another procedure proposed in the literature is one in which neighboring cells send each other periodically an indication of their channel utilization. By predicting ahead, a given cell can determine the chance of a newly admitted call being denied service in a neighboring cell if it is subsequently handed off. If that probability turns out to be above a given threshold, it is better to deny service to the new call in the first place. Calculations indicate that this strategy provides an improvement over the guard-channel scheme, but it does require periodic communication between cells. Other simple scheme is that of buffering handoff calls up to some maximum time if no channel is initially available. The handoff dropping probability does of course reduce as a result, at the cost of a delay in continuing service. If this delay is not too high, it may be acceptable to the participants in an ongoing call.[18] In this paper the Guard-the channel scheme has been studied.

In the related work session, examples of algorithms in the literature have been studied. In the next session, the proposed methodology has been introduced. Then in **implementation and evaluation** Session, performance metrics for these algorithms are calculated and optimal algorithm has been found between them.

## 2   Background and related Works

Handover algorithms are classified from different view. To reduce the handover latency, two categories of protocols have been proposed. One focuses on the change in network architecture such as HMIP and IDMP. The other focuses on the mechanism to reduce latency by MN and AR(Access Router) themselves, hence change in design, such as fast handover. In this paper, Examples of each class of handover is considered so that change in design or architecture is evident.

### 2.1 Change in design

In design change process, the characteristics of MIPv6 (Mobile Internet Protocol version 6) are implemented to improve the efficiency parameters. Some important protocols as fast handover enhanced fast handover and seamless MIPv6 will be discussed as follows.

#### 2.1.1 Fast Handover Protocol

The protocol enables an MN to do movement detection and create nCoA(New CoA), by providing the new access point and the associated subnet prefix information when the MN is still connected to its current subnet[19]. Unlike in FMIPv6 algorithms in MIPv6, L2 handover should be done before L3 handover. Handover in layer 2 includes: channel scanning, association and authentication.[20]

In FMIPv6 to prevent the packet loss, a bidirectional tunnel between PAR and NAR is established. the binding updates to the HA and CN(Correspondent Node) are performed after the time point when the MN is IP-capable on the new subnet link.[3] Because of this, the MN communicates with the CN directly via the NAR, before completing the BU, using this tunnel in a very late time. Figure 1 shows the messages exchanged during FMIPv6.

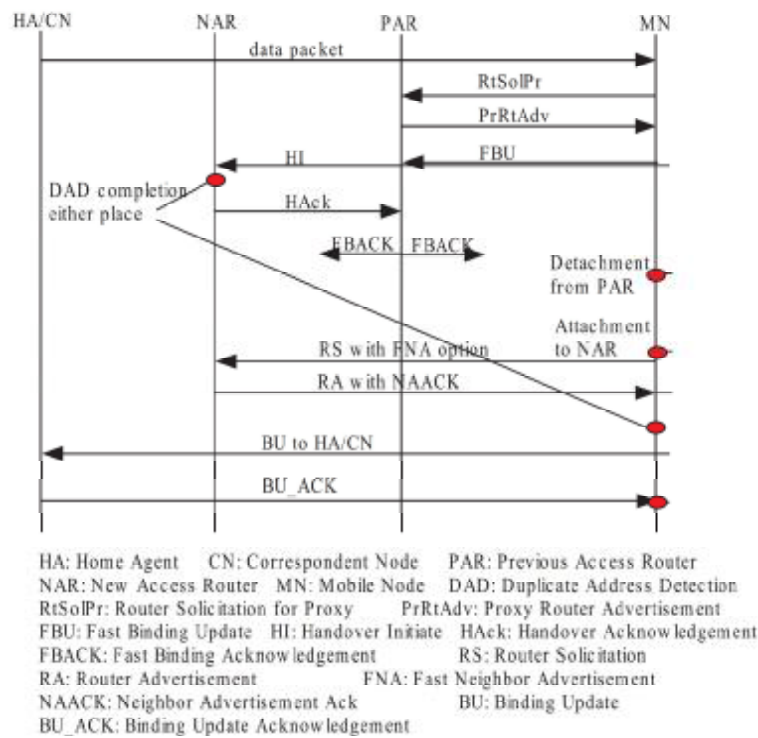

Figure 1: messages exchanged during FMIPv6[3]

#### 2.1.2 Enhanced Fast Handover Protocol

In EFMIPv6, LI has stated that, unlike the FMIPv6 the nCoA generation and DAD procedure can be performed before handover starts. At the same time, that when nCoA is informed to PAR, the handover to the new access point will definitely happen. Therefore, It is known that the binding update to the HA/CN can be performed at the time point when the new CoA is known by PAR. Also It has allowed that new AR construct a new CoA, perform DAD for the MN and store this new CoA to the nCoA table when anticipating that a handover for an MN is about to happen. At the same time, to reduce the registration latency in the binding update, the binding update to the HA/CN will be performed after the PAR knows the nCoA.[2] To describe the optimized scheme clearly, the detailed timing graph for the enhanced scheme is provided in Figure 2.

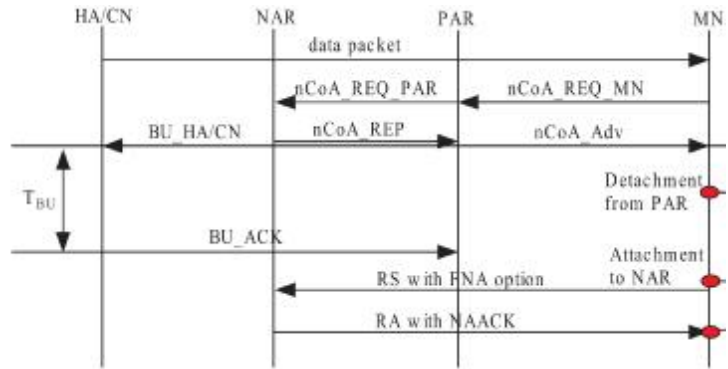

Fig.2: messages exchanged during EFMIPv6[3]

### 2.1.3 Seamless Mobile IPv6 Protocol

SMIPv6 makes use of users' mobility patterns to predict the cell where the next handover will occur. Based on this knowledge, the protocol updates all its CNs with its new address before leaving its current network and entering a new one. Furthermore, using layer 2 information, SMIPv6 is able to predict the exact time the handover will occur. Using its mobility pattern, a mobile node will send update messages to its correspondent nodes only when a change of network is in sight. Normally, these updates occur at regular intervals. SMIPv6's mobility management model is divided into two components: a mobility pattern learning module implemented in each mobile node and a mobility management protocol executed by all entities in the network. The L3 handover is performed upon the reception of a layer 2 trigger. The trigger contains identification information about the new access point. Based on this identifier, a mobile node can verify if this AP(Access Point) is part of its mobility profile. If it is, the NCoA based on the sub network's prefix is created without waiting for the RAs to be sent by the AR. Upon the completion of the address creation phase, the MN sends BUs containing its NCoA to all its CNs as well as to its HA. Then in this algorithm delay of RtSolPr and PrRtAdv messages exchange and delay of BU are deleted. Fig. 3 shows the messages exchanged during SMIPv6 handover.[21]

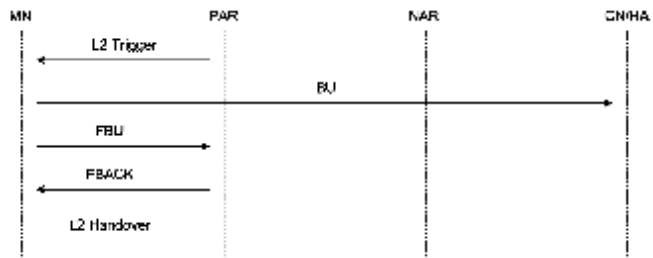

Figure 3 : messages exchanged during SMIPv6[21]

## 2.2 Change in architecture

In architecture change process, one or more entities to improve performance are added to the existing architecture. For example in HMIPv6, one or more MAP(Mobility Anchor Point) are added to the network architecture or in[2] functional network entity, called the handover coordinator (HC), to the IP core to be shared and utilized by the internetworking heterogeneous wireless networks (i.e. both source and target networks) in a PMIPv6 micro-mobility domain.

### 2.2.1 Hierarchical MIPv6 protocol

This method is design for handover delay problem when the HA or CN is located geographically far away from the MN and when a mobile node moves in a small coverage area (micro-mobility).[10] Authenticating binding updates requires approximately 1.5 round-trip times between the mobile node and each correspondent node. In addition, one round-trip time is needed to update the Home Agent; this can be done simultaneously while updating correspondent nodes. For these reasons a new Mobile IPv6 node, called the Mobility Anchor Point, is used and can be located at any level in a hierarchical network of routers, including the AR. The MAP will limit the amount of Mobile IPv6 signaling outside the local domain  The introduction of the MAP provides a solution to the issues outlined earlier in the following way:

 - The mobile node sends Binding Updates to the local MAP rather than the HA (which is typically further away) and CNs.

 - Only one Binding Update message needs to be transmitted by the MN before traffic from the HA and all CNs is re-routed to its new location.  This is independent of the number of CNs that the MN is communicating with.[7]

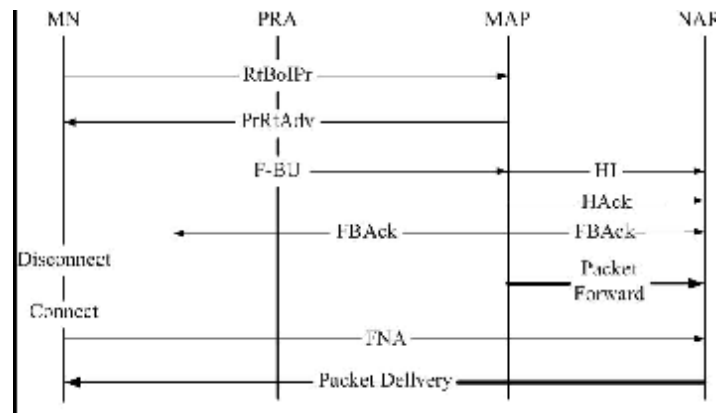

Figure 4 : messages exchanged during HMIPv6 [22]

## 3 Proposed methodology

The proposed methodology to choose the best and proper protocol in different situations includes four steps. It should be noted that voice packet as an example, is used in data analyzing.

### 3.1 Determine handover class

In the first step the class of studied handover algorithms should be determines. In the proposed methodology, handover algorithms occurred in the network-layer that can be run horizontally are compared. Determining the time of connection to the new access point, is important to determining the number of channels used in the algorithm. Determining the geographic scope for the studied algorithms is important to feasibility of change in design or architecture.

### 3.2 The performance metrics calculation

After determine the class of each algorithm, in the second step, according to the topology used in Figure 5, the delay of each step should be calculated. Processing delay of a node n, is assumed equal to T. All delays on wired links hold value f except for link (N1, N2) which holds value F. This link represents both local and global mobility and in HMIPv6 study, determine domain. Each radio Link will have a delay equal to d. L2 Handover delays hold a value equal to h. It is necessary to note that, except processing delay and propagation delay, other delays are ignored. But other delay scan be easily calculated or based on Cisco recommends[23], using worst case in design. Due to the importance of DAD delay, in proposed methodology, this delay is calculated separately. This delay in the worst case that referred in MIPv6 reference algorithm[24], is intended *D= 1 s* and is add to total signaling delay of handover algorithm that is not adjusted or deleted on them. It can be seen the calculating details of performance parameters for the mentioned protocols in the following. To become more transparent, the results of the MIPv6 reference algorithm also have been studied.

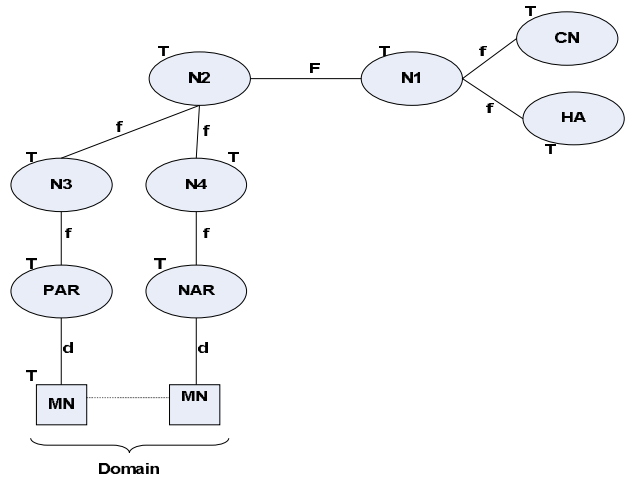

Fig.5: Proposed topology for evaluating handover algorithms performance

#### 3.2.1 Case 1: MIPv6 handover

Fig. 6 shows messages exchanged during an MIPv6 handover. Table 1 points out the chronological details of messages exchanged as well as the analytical delay found for each event. The last packet through the PAR was received at t = T. The first packet through the NAR was received at *t = 36T+22f+6d+h+2F*. Hence, the total handover delay amounts to: *t = 35T+22f+6d+h+2F*

From the moment where the MN initiates the handover to when the CN sends its packets to the new NCoA, packets sent to the previous CoA are lost. The exact number of packets lost can be calculated using the following formula: *(35T+22f+6d+h+2F)* Throughput*.

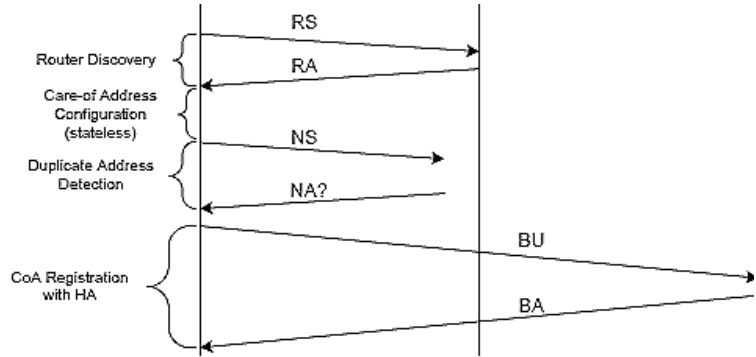

Figure 6 : MIPv6 signaling[25]

The signalization latency starts precisely when the mobile node receives the RA and ends when the BU is received by the MN's correspondent node. Thus, the total value of the signalization delay is equal to: $t = 30T+19f+5d+h+F+D$.

Table 1:. Chronological details of an MIPv6 handover

| Delay | Event | Time |
|---|---|---|
| T | L2 Trigger | $t = 0$ |
| $6T+4f+d$ | RS | $t = T$ |
| $6T+4f+d$ | RA | $t = 6T+4f+d$ |
| $6T+4f+d$ | NS | $t = 12T+8f+2d$ |
| $6T+4f+d$ | NA | $t = 18T+12f+3d$ |
| H | L2 Handover | $t = 24T+16f+4d$ |
| $3f+F+d+6T$ | BUs sent to HA/CN | $t = 24T+16f+4d+h$ |
| $3f+F+d+6T$ | Packets sent by CNs@NCOA | $t = 30T+19f+5d+h+F$ |
|  | Packets sent by CNs are received | $t = 36T+22f+6d+h+2F$ |

### 3.2.2 Case 2: FMIPv6 handover

Figure 1 shows the messages exchanged during an FMIPv6 handover. Table 2 points out the chronological details of messages exchanged as well as the analytical latency found for each event. The last packet through the PAR was received at $t = 4d + 8f + 18T$. The first packet through the NAR was received at $t = max\ t = max\ (6d +8f + h + 22T, 12f +4d + 23T)$. Hence, the total handover delay is given by: $Max\ (2d + h + 4T, 4 + 5T)$

No packets are lost since the PAR starts rerouting packets toward the NAR before proceeding with the handover .All packets received in the meantime, that is, before the L2 handover is performed, are stored in a buffer thus ensuring that no packets are lost. Following the reception of the FNA, all packets are sent to the MN. Although packet losses are null, the signalization delay is quite high. The L2 trigger is only received by the MN at time and the CN and HA receive their respective BUs at $t =11f +7d + F + h + 28T$. Thus, the signalization delay is equal to: $14f +6d + 2F + h + 30T+D$

### 3.2.3 Case 3: SMIPv6 handover

Figure 3 shows the messages exchanged during an SMIPv6 handover. Table 3 presents the chronological details of messages exchanged as well as the analytical delay found for each event. The last packet going through the PAR is received at $t = 2d + 5T$. The first packet passing through the PAR is received at t = min = *min(2d + 4f + 9T, 2d + 6f +2F + 12T)*. Hence, the handover delay is equal to: *4f + 4T*

There are no packets lost since the PAR reroutes packets through the NAR before performing the actual handover. Indeed, the MN joins the new network before packets sent by the CNs or rerouted by the PAR reach the new network. the first rerouted packet arrive at *4f + 3d + 6T* and that the MN joins the new *network at 2d + h + 5T* Thus, if we subtract the time the rerouted packets arrive from the time the MN reaches its new network, we get 4f _ h + T, a positive value since h is near 0 (L2 handover delay) and T is relatively small. The signalization delay equal to: 3f + F + d + 5T

Table 2: Chronological details of a FMIPv6 handover

| Delay | Event | Time |
|---|---|---|
| *T* | L2 Trigger | *t = 0* |
| *d+2T* | RtSolPr | *t = T* |
| *d+2T* | PrRtAdv | *t=d+2T* |
| *d+2T* | FBU | *t =2d + 4T* |
| *4f + 5T* | HI | *t =3d + 6T* |
| *4f + 5T* | HACK | *t =3d +4f + 11T* |
| *d+2T* | FBACK | *t =3d +8f + 16T* |
| *4f + 5T* | Packets are rerouted through | *t =3d +8f + 16T* |
| *h* | L2 Handover | *t =4d +8f + 18T* |
| *d+2T* | FNA | *t =4d +8f + h + 18T* |
| *d+2T* | FNA -ACK | *t =5d +8f + h + 20T* |
| *3f + F + d + 6T* | BUs sent to HA/CN | *t =6d +8f + h + 22T* |
| *d+2T* | PAR sends packets to MN | *t = max (5d +8f + h + 20T, 12f +3d + 21T)* |
|  | Packets are received by MN | *t = max (6d +8f + h + 22T, 12f +4d + 23T)* |
|  | BUs are received by CNs | *11f +7d + F + h + 28T* |
|  | Bus-ACK are received by MNs | *t =14f +8d + 2F + h +* |

Table 3: Chronological details of a SMIPv6 handover

| Delay | Event | Time |
|---|---|---|
| T | L2 Trigger | *t = 0* |
| *d + 2T* | FBU | *t = T* |
| *3f + F + d + 6T* | BU | *t = T* |
| *d + 2T* | FBACK | *t = d + 3T* |
| *4f + d + 6T* | Rerouting of packets | *t = d + 3T* |
| *h* | L2 Handover | *t =2d + 5T* |
| *3f + F + d + 6T* | Packets sent by CNs@NCOA | *t = d +3f + F + 6T* |
|  | Rerouted packets are received | *t = 4f +2d + 9T* |
|  | Packets sent by CNs are received | *t = 6f +2F +2d + 12T* |

### 3.2.4 Case 4: EFMIPv6 handover

Figure 2 shows the messages exchanged during an EFMIPv6 handover. Table 4 presents the chronological details of messages exchanged as well as the analytical delay found for each event. Like as fast handover No packets are lost, then the handover delay is equal to: *max (3f+h+F+5T,4f+3T )*.The signalization delay equal to:*3d+11f+21T+F+h*

### 3.2.5 Case 5: HMIPv6 handover

Figure 4 shows the messages exchanged during a HMIPv6 handover. Table 5 presents the chronological details of messages exchanged as well as the analytical delay found for each event. Like as fast handover No packets are lost, then the handover delay is equal to: *max (2d + h + 6T, 2f +d+ 5T)*. The signalization delay equal to: *10f +3d + h + 19T+D*

Table 4: Chronological details of a EFMIPv6 handover

| Delay | Event | Time |
|---|---|---|
| T | L2 Trigger | t = 0 |
| d + 2T | nCoA-REQ-MN | t = T |
| 4f + 5T | nCoA-REQ- PAR | t = d + 2T |
| 4f + 5T | nCoA-REP | t =d+4f+7T |
| 3f + F + 5T | BUs sent to HA/CN | t =d+8f+12T |
| d+2T | nCoA-Adv | t =d+8f+12T |
| 4f + 5T | Packets are rerouted through PAR | t =d+8f+12T |
| 3f + F + 5T | BU_ACK | t =d+11f+17T+F |
| H | L2 Handover | t =d+11f+17T+F |
| d+2T | FNA | t =d+11f+17T+F+h |
|  | Rerouted packets are received | t =2d+12f+17T |
|  | Packets are received by MN | t=max(2d+11f+19T+F+h, 2d+12f+17T) |
| d+2T | NAACK | t =2d+11f+19T+F+h |
|  | NAACKs are received by MN | t =3d+11f+21T+F+h |

Table 5: Chronological details of a FHMIPv6 handover

| Delay | Event | Time |
|---|---|---|
| T | L2 Trigger | t = 0 |
| d+2T | RtSolPr | t = T |
| d+2T | PrRtAdv | t = d + 2T |
| d+2T | FBU | t =2d + 4T |
| 4f+5T | HI | t =3d + 6T |
| 4f+5T | HACK | t =3d +4f + 11T |
| d+2T | FBACK | t =3d +8f + 16T |
| 4f + 5T | Packets are rerouted through PAR | t =3d +8f + 16T |
| h | L2 Handover | t =4d +8f + 18T |
| d+2T | FNA | t =4d +8f + h + 18T |
| d+2T | FNA -ACK | t =5d +8f + h + 20T |
| 2f +d+T | BUs sent to MAP | t =6d +8f + h + 22T |
|  | PAR sends packets to MN | t = max (5d +8f + h + 20T, 12f +3d + 21T) |
|  | Packets are received by MN | t = max (6d +8f + h + 22T, 12f +4d + 23T) |
|  | BUs are received by MAP | t =10f +7d + h + 23T |
|  | Bus-ACK are received by MN | t =12f +8d + h + 24T |

After calculating values of Packet loss ،Handover Delay and Signaling Delay, using available formulas,[18] we can calculate Call blocking and Handover blocking probability.

### 3.3 Weighting the metrics based on AHP algorithm

Performance metrics in each method is obtained, the weight of these metrics should be allocated, till can use these metrics in MCDM methods. AHP, fuzzy AHP, fuzzy TOPSIS, TOPSIS methods respectively, are as most efficient MCDM Compensatory methods.[26]

The work of selecting the appropriate handover method in the literature[27, 28], AHP technique as a method of weighting the quantitative and qualitative criteria are considered.

### 3.4 The appropriate method according to TOPSIS

According to the literature[5] in the Fourth step, using TOPSIS algorithm among the various available handover methods, appropriate method is selected.

## 4 Implementation and evaluation

In this section the performance of FMIPv6 ،EFMIPv6 ،SMIPv6 and HMIPv6 will be evaluated according to the described methods in the previous section.

### 4.1 Handover class

Class of each method determine in table 6. In SMIPv6 protocol, to preparation and installation mobility pattern learning module on each node and planning and implementation of the mobility management protocol to the project cost will be added. In HMIPv6 protocol, to add a MAP, the cost will be added to the project.

Table 6: Classifying studied algorithms

| Change in design/ architecture | support micro/macro mobility | Hard/Soft handover* | Algorithm Class |
|---|---|---|---|
| --------------------------- | macro mobilitysupport | Hard | MIPv6 |
| Change in design | support macro mobility | Soft | FMIPv6 |
| Change in design | support macro mobility | Soft | EFMIPv6 |
| Change in design | support macro mobility | Soft | **SMIPv6 |
| Change in architecture | support micro mobility | Soft | ***HMIPv6 |

*The algorithms are implemented as soft, only half of the channels are available.

### 4.2 The performance metrics calculation

To calculating performance metrics, the following conditions are considered:

Speed of mobile node: 60 km/h, average call holding time is 300 sec and cell radius is r = 10 km. There are ten channels in each cell that three channels are considered as guard channels. Using the above values, can be calculate Call blocking probability and Handover blocking probability. Also, we have:

Propagation speed on the wireless link is equal to $2*10^8$ m/s. Propagation speed on the wired link is equal to $3*10^8$ m/s. the length of wireless link d= 500 m, the length of f wired link is f=35m and the length of F wired link is F= 2 km. Then the propagation delays on above link are 2.5 μsec ،0.12 μsec and 6.7 μsec respectively. Given ADPCM, G.726as a coder, the processing delay at each node in best case is equal to T= 2.5 ms .[23] The cost for reference algorithm is considered as 1000. The results of the algorithms are given in Table 7 and 8.

Table 7: The results of evaluating algorithms as parametric

| Algorithm | Packet lost | Handover Delay | Call blocking probabilit | Handover blocking probabilit | Signaling Delay | Price |
|---|---|---|---|---|---|---|
| MIPv6 | $35T+22f+6d+h+2F$ | $12T+6f+2d+h+2F$ | $1.82*10^{-3}$ | $6.74*10^{-11}$ | $30T+19f+5d+h+2F+D$ | 1000 |
| FMIPv6 | 0 | $\max(2d+h+6T, 4f+7T+d)$ | 0.56 | $2.5*10^{-5}$ | $14f+6d+2F+h+30T+D$ | 1000 |
| SMIPv6 | 0 | $4f+4T$ | 0.56 | $2.5*10^{-5}$ | $3f+F+d+5T+D$ | 1500 |
| EFMIPv6 | 0 | $\max(3f+h+F+5T, 4f+3T)$ | 0.56 | $2.5*10^{-5}$ | $3d+11f+21T+F+h$ | 1000 |
| HMIPv6 | 0 | $\max(2d+h+6T, 2f+d+5T)$ | 0.56 | $2.5*10^{-5}$ | $10f+3d+h+19T+D$ | 1500 |

Table 8: The results of evaluating algorithms as numerical

| Algorithm | Packet lost | Handover Delay | Call blocking probability | Handover blocking probability | Signaling Delay | Price |
|---|---|---|---|---|---|---|
| MIPv6 | 0.08753104 | 0.0300191 | $1.82*10^{-3}$ | $6.74*10^{-11}$ | 1.0750281 | 1000 |
| FMIPv6 | 0.00000001 | 0.0175029 | 0.56 | $2.5*10^{-5}$ | 1.0750300 | 1000 |
| SMIPv6 | 0.00000001 | 0.0100004 | 0.56 | $2.5*10^{-5}$ | 1.0125095 | 1500 |
| EFMIPv6 | 0.00000001 | 0.0125070 | 0.56 | $2.5*10^{-5}$ | 0.0525155 | 1000 |
| HMIPv6 | 0.00000001 | 0.015005 | 0.56 | $2.5*10^{-5}$ | 1.0475087 | 1500 |

### 4.3 Weighting the metrics based on AHP

To weight to metrics, using experts' opinion. Finally, weight of each metric with respect to the output of the software is as follows:

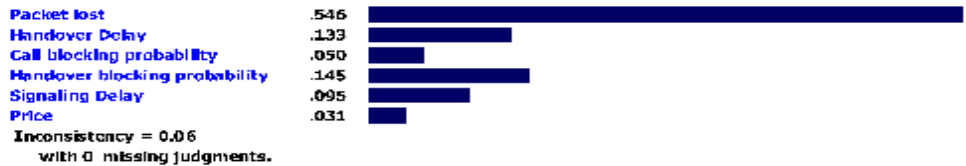

Fig.7: the weight of each metrics according to the expert choice software

### 4.4 The appropriate method according to TOPSIS

Decision matrix to select the optimal Handover algorithm, after calculating all the types of metrics shown in table 9.

Table 9: Decision matrix to select the optimal Handover algorithm

| Algorithm | Packet lost | Handover Delay | Call blocking probability | Handover blocking probability | Signaling Delay | Price |
|---|---|---|---|---|---|---|
| MIPv6 | 0.00011854 | 0.00004912 | $1.82*10^{-3}$ | $6.74*10^{-11}$ | 1.00010318 | 1000 |
| FMIPv6 | 0 | 0.000015 | 0.56 | $2.5*10^{-5}$ | 1.00010508 | 1000 |
| SMIPv6 | 0 | 0.00001048 | 0.56 | $2.5*10^{-5}$ | 1.00002206 | 1500 |
| EFMIPv6 | 0 | 0.00001956 | 0.56 | $2.5*10^{-5}$ | 0.00006802 | 1000 |
| HMIPv6 | 0 | 0.000015 | 0.56 | $2.5*10^{-5}$ | 1.0000612 | 1500 |

Finally, the rating options are as follows:

1. EFMIPv6
2. SMIPv6
3. HMIPv6
4. FMIPv6
5. MIPv6

# 5 Conclusion

The handover process happens when the MN moves from one access medium to another, and it should accomplish three operations: movement detection, new CoA configuration, and BU. During handover period, the MN is unable to send or receive packets as usual. The length of this period which is called handover latency is very critical for the delay-sensitive and real-time services. To reduce the handover latency and increase its efficiency several methods have been proposed in the literature. In this paper, a methodology for choosing the appropriate algorithm between the existing methods is presented. It was clarified that BU and DAD signaling are critical points of handover algorithms then methods that try to improve this point, are successful in improving the overall effectiveness of Handover.

As expected, EFMIPv6 protocol is the best selection, because of eliminate DAD delay and reduce the delay of BU. The cost of the SMIPv6 algorithm is increased and the time required for BU signaling effectively reduced and time needed to exchange RtSolPr and PrRtAdv messages are deleted. Normally, in practical, Algorithms that have changed in design or architecture should be examined separately. In HMIPv6 algorithm, when the mobile node moves within a domain, If the change in topology in HMIPv6 and MAP or MAPs is/are adding, increase the cost of this algorithm should also be considered. Despite packet loss in the algorithms that use of hard handover, when traffic is low sensitivity to packet loss, weight of packet loss parameter in the AHP algorithm is reduced and due to the efficient use of bandwidth in these algorithms, their use is preferred. For evidence result, algorithms have been selected that, have obvious difference. But in methods that in which change in design or architecture are complex or similar, using proposed methodology is very effective.